\documentclass{article}

\usepackage[preprint, nonatbib]{nips_2018}

\usepackage[utf8]{inputenc} 
\usepackage[T1]{fontenc}    
\usepackage{hyperref}       
\usepackage{url}            
\usepackage{booktabs}       
\usepackage{amsfonts}       
\usepackage{nicefrac}       
\usepackage{microtype}      

\usepackage{comment}
\usepackage{multirow}
\usepackage{subcaption}
\usepackage{graphicx}
\graphicspath{ {Images/} }

\title{Predicting Onset of Dementia in Parkinson's Disease Patients}

\author{
  Dhruv Agarwal \\
  Department of Computer Science\\
  University of Illinois at Urbana-Champaign\\
  Champaign, IL 61820 \\
  \texttt{dhruva2@illinois.edu} \\
   \And
  Abhishek Srivastava \\
  Department of Computer Science\\
  University of Illinois at Urbana-Champaign\\
  Champaign, IL 61820 \\
  \texttt{as29@illinois.edu} \\
  \And
  Edward W Huang \\
  Department of Computer Science\\
  University of Illinois at Urbana-Champaign\\
  Champaign, IL 61820 \\
  \texttt{ewhuang3@illinois.edu} \\
}

\begin{document}

\maketitle

\begin{abstract}
Alzheimer's disease (AD) and Parkinson's disease (PD) are the two most common neurodegenerative disorders in humans. Because a significant percentage of patients have clinical and pathological features of both diseases, it has been hypothesized that the patho-cascades of the two diseases overlap. Despite this evidence, these two diseases are rarely studied in a joint manner. In this paper, we utilize clinical, imaging, genetic, and biospecimen features to cluster AD and PD patients into the same feature space. By training a machine learning classifier on the combined feature space, we predict the disease stage of patients two years after their baseline visits. We observed a considerable improvement in the prediction accuracy of Parkinson's dementia patients due to combined training on Alzheimer's and Parkinson's patients, thereby affirming the claim that these two diseases can be jointly studied.

\end{abstract}

\section{Introduction}

Alzheimer’s disease (AD) is an irreversible, progressive neurological disease that gradually deteriorates cognitive abilities~\cite{Irvine2008}. It is characterized by progressive impairment in memory, judgment, language, and orientation to physical surroundings. The primary pathological features of AD are neuronal loss and the accumulation of extracellular plaques containing amyloid $\beta$ (A$\beta$) and neurofibrillary tangles (NFT) containing tau. Parkinson's disease (PD) is a similarly irreversible, progressive neurological disease that impairs movement. PD manifests as a movement disorder and a distinct form of cognitive impairment with typical pathological characteristics of the accumulation of $\alpha$-synuclein in multiple brain regions. 

Though AD and PD are clinically distinct entities, one can find pathological associations between them. For instance, A$\beta$, an important hallmark of AD pathology, has been reported to be present in some PD patients \cite{kempster2010relationships}, while there have been indications that Lewy body deposition exists even in AD patients \cite{compta2011lewy}. The association might also extend to genes such as APOE and MAPT genes which have been reported to be linked to the presence of A$\beta$ aggregates \cite{polvikoski1995apolipoprotein} and increased tau protein expression. 

An important application of the hypothesis of overlap between AD and PD has been studying cognitive decline in PD patients. Previous works report association of lower levels of cerebrospinal fluid (CSF) amyloid $\beta$ with cognitive decline in PD patients~\cite{shaw2009cerebrospinal}. Also, many imaging-based measures have been reported to be associated with cognitive performance in PD. A baseline AD pattern of brain atrophy, quantified using the Spatial Pattern of Atrophy for Recognition of AD (SPARE-AD) score, predicted long-term global cognitive decline in non-demented PD patients~\cite{weintraub2011alzheimer}.

Despite the discovery of such correlations, AD and PD communities remain distinct. This motivates an aggregated, data-driven study to explore correlation between the two diseases, which could potentially lead to clearer understanding and possible development of therapeutic targets for both ailments. The main contribution of this paper is combining multiple modalities of two different data sets of ADNI and PPMI patients and consequently predicting the onset of Dementia in PD patients using it. To the best of our knowledge, this is the first work exploiting the overlap of these two diseases to improve prediction accuracy of dementia in PD patients.
\section{Methodology}

\subsection{Problem Statement}

The aim is to predict a given patient’s cognitive status (Normal CI/MCI/Dementia) after 24 months using his/her baseline features. Due to high clinical importance of prediction of onset of Dementia in PD patients, the primary goal is to improve on state-of-the-art methods for the same. 

\subsection{Data}

Alzheimer’s Disease Neuroimaging Initiative (ADNI) database was used to fetch multimodal data for AD patients. ADNI-2 participants meeting the criteria for single-domain or multidomain amnestic MCI were selected, as in prior work \cite{mathotaarachchi2017identifying}. The availability of all desired features and diagnosis after $24$ months led to a total  of $355$ patients. Parkinson’s Progression Markers Initiative (PPMI) database was used to fetch multimodal data for PD patients. Montreal Cognitive Assessment (MoCA) scores~\cite{nasreddine2005montreal} were used for assignment of Dementia labels for PD patients. MoCA is a widely used screening assessment for detecting cognitive impairment ~\cite{Dalrymple-Alford1717}~\cite{Hoops1738}. Based on the availability of MoCA scores, and other required features, we were reduced to $142$ patients.

\subsection{Features}

\begin{figure}[th!]
    \begin{subfigure}{0.5\columnwidth}
    \includegraphics[width=\linewidth, height=5cm]{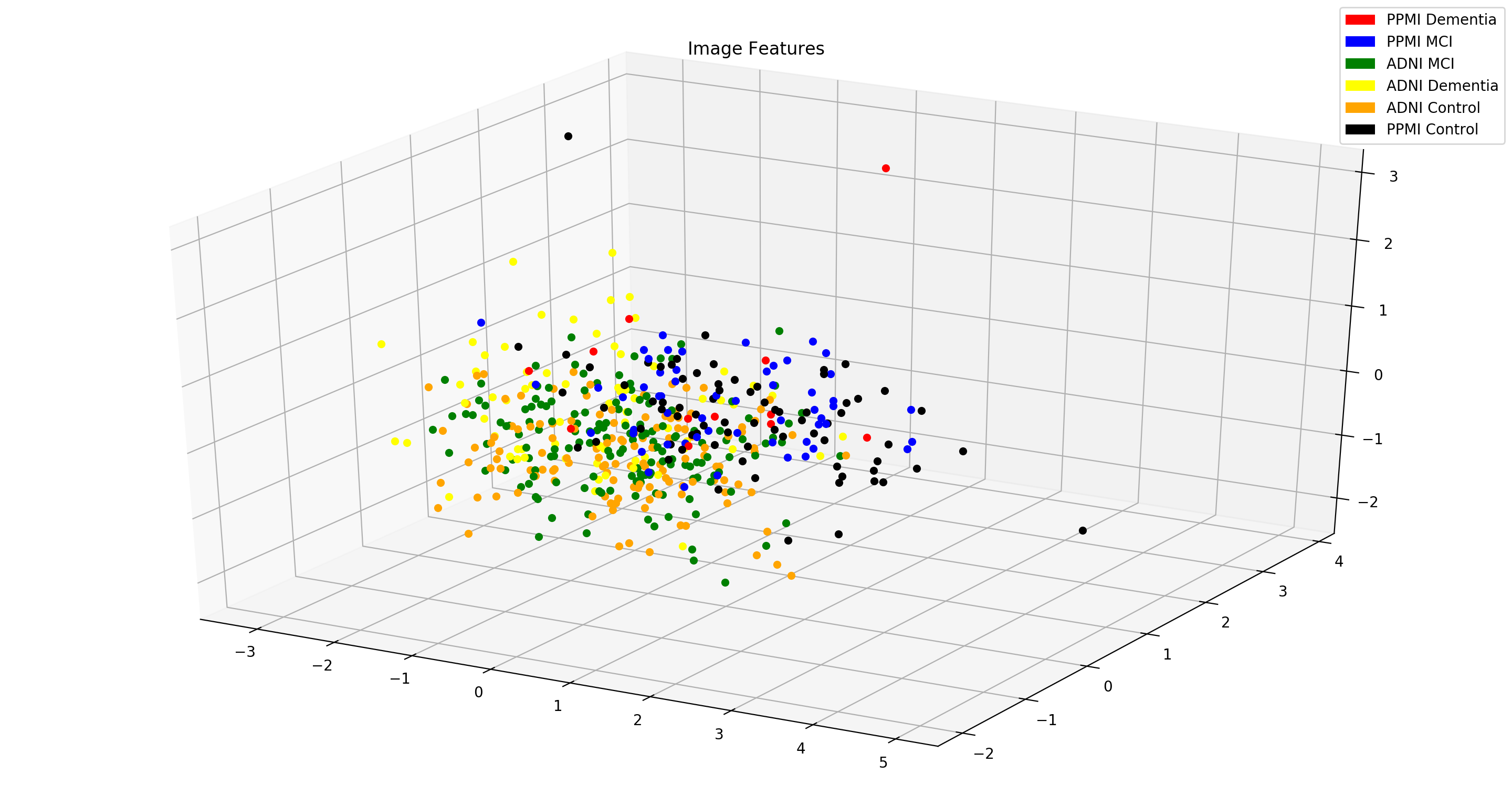}
    \caption{}
    \label{fig:feature_desc_1}
    \end{subfigure}
    \begin{subfigure}{0.5\columnwidth}
    \includegraphics[width=\linewidth, height=5cm]{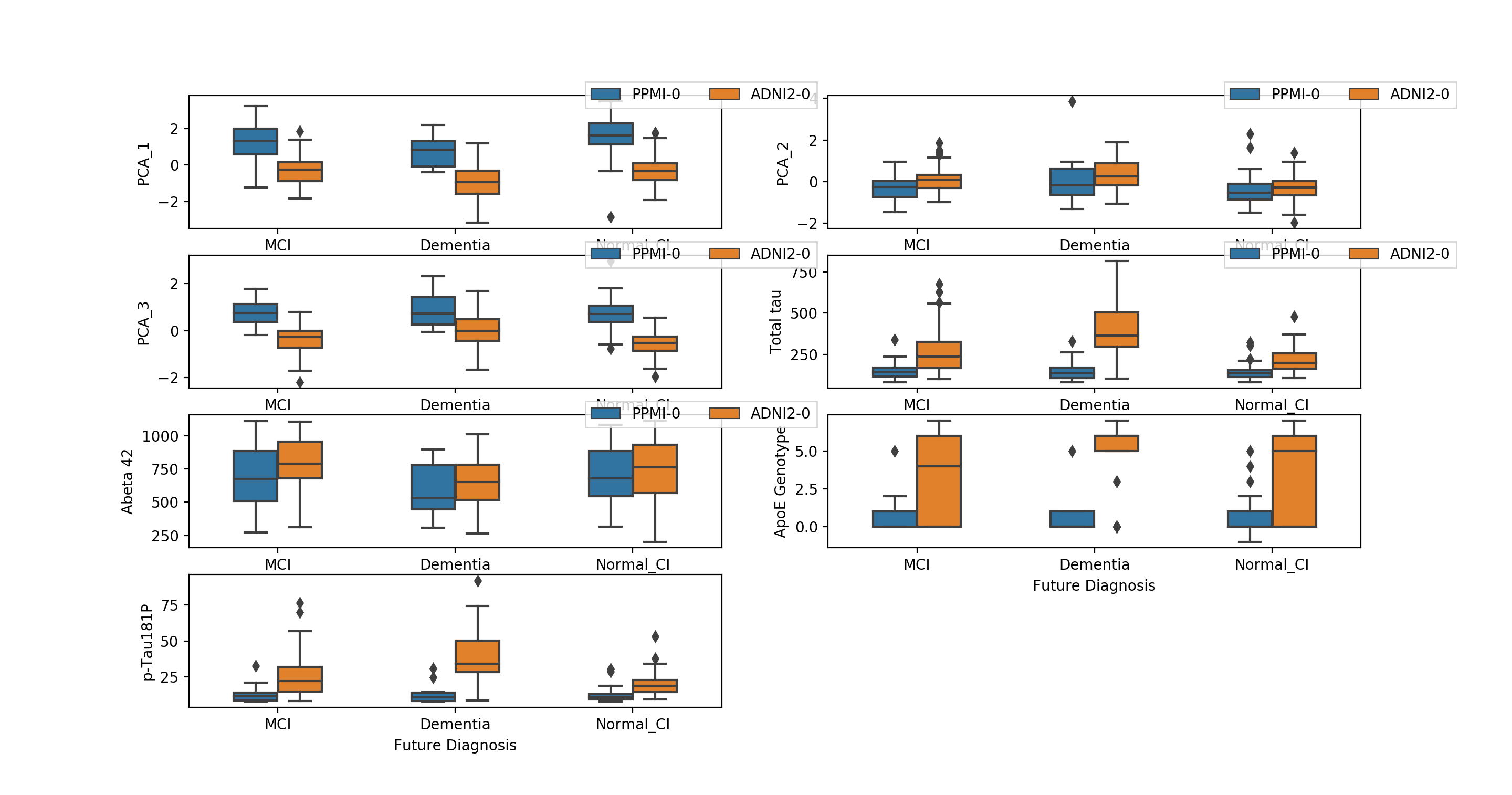}
    \caption{}
    \label{fig:feature_desc_2}
    \end{subfigure}
\caption{a) Reduced Image feature space comprising of all patients (AD/PD) b) All feature overlap between AD and PD patients}
\label{fig:feature_desc}
\end{figure}

The following features were selected for the purpose of this classification task. These features have been found to drive dementia prediction~\cite{berlyand2016}. 

\subsubsection{Imaging Features}

FreeSurferV5.1 software~\cite{fischl2012freesurfer}, which can analyze and visualize MRIs, was used to obtain pre-processed image features. This allowed us to map ADNI and PPMI patients into a unified image feature space. Principal Component Analysis (PCA)~\cite{pearson1901liii} was used to reduce the feature space to three dimensions, as illustrated in Fig ~\ref{fig:feature_desc_1}. Alternative methods such as independent component analysis and t-sne did not perform as well as PCA on this data. Note that each data point is labeled by its cognitive label after $24$ months in addition to its dataset of origin. We observe that data points in the ADNI and PPMI datasets that share common cognitive statuses are in close proximity with one another. 

\subsubsection{Genetic Features}

Apolipoprotein E4 (ApoE4), which is the largest known genetic risk factor for late-onset sporadic AD in a variety of ethnic groups~\cite{corder1993gene, sadigh2012association} was chosen as the only genetic feature. ApoE4 has also been shown to be associated with cognitive decline in Parkinson's disease~\cite{aarsland2017cognitive}.

\subsubsection{Biospecimen Features}

For biospecimen features, cerebrospinal fluid (CSF) total tau, phosphorylated tau 181 protein (p-tau181), and CSF A$\beta 42$ levels were selected. Elevated CSF total tau and p-tau181 concentrations have shown to be associated with neuro degenerative changes in early AD~\cite{thomann2009association, shaw2009cerebrospinal, younkin1998role}. Each of these biomarkers also have strong prognostic and diagnostic potential in early-stage PD~\cite{kang2013association}. CSF alpha-synuclein levels were also considered as a potential feature but could not be used due to unavailability of data for ADNI-2/GO participants~\cite{irwin2017neuropathological}. 

One can observe significant overlaps between feature values of clustered ADNI and PPMI patients in Fig ~\ref{fig:feature_desc_2}, suggesting potential for accurate predictions for both kinds of patients.

\subsection{Classifier}
Using the features at baseline (gender, three imaging features, ApoE4 genotype, and the three biospecimen features), a Random Forest classifier (500 trees with max features parameter as 0.6) was trained to predict a patient's cognitive disease label two years after the baseline. Preliminary testing showed Random Forest outperforming other commonly used classifiers such as logistic regression, support vector machine, multi-layer perceptron etc. This corroborates with the fact that Random forests are inherently suited for multi-class problems and work well in combination with numerical, categorical features. Five-fold cross-validation, splitting the folds on AD and PD patients independently, and recombining them for the training to avoid imbalances between the two diseases, was used. Thus, each fold contained equal proportions of AD and PD patients.

\section{Results}

Table~\ref{tab:results} summarizes the results for the aforementioned classifier, testing different train-test set combinations. Note that even though we list accuracy as a performance parameter, it was not used as a measure of comparison because it can misinterpret the performance when the sample has imbalanced class probabilities~\cite{mathotaarachchi2017identifying}. 

\begin{table}[h!]
    \caption{Summary of results}
    \label{tab:results}
    \centering
    \resizebox{\textwidth}{!}{
    \begin{tabular}{ccccccccc}
    \toprule
    Training set & Test set & AUC-Dementia & AUC-MCI & AUC-CI & Accuracy & Precision & Recall\\ [0.5ex] 
    \midrule
    ADNI+PPMI & ADNI+PPMI &  0.88 & 0.64 & 0.72 & 0.53 & 0.57 & 0.53\\
    ADNI+PPMI & PPMI  & \textbf{0.95} & 0.53  & 0.62 & 0.57 & 0.33 & 0.36 \\
    ADNI+PPMI & ADNI &  0.91 & 0.68  & 0.77 & 0.52 & 0.55 & 0.53 \\
    PPMI & PPMI &  \textbf{0.61} & 0.42  & 0.65 & 0.48 & 0.27 & 0.32 \\
    ADNI & ADNI &   0.89 & 0.67  & 0.77 & 0.52 & 0.56 & 0.55 \\
    \bottomrule
\end{tabular}}
\end{table}


\begin{figure}[th!]
    \begin{subfigure}{0.33\columnwidth}
    \includegraphics[width=\linewidth, height=4cm]{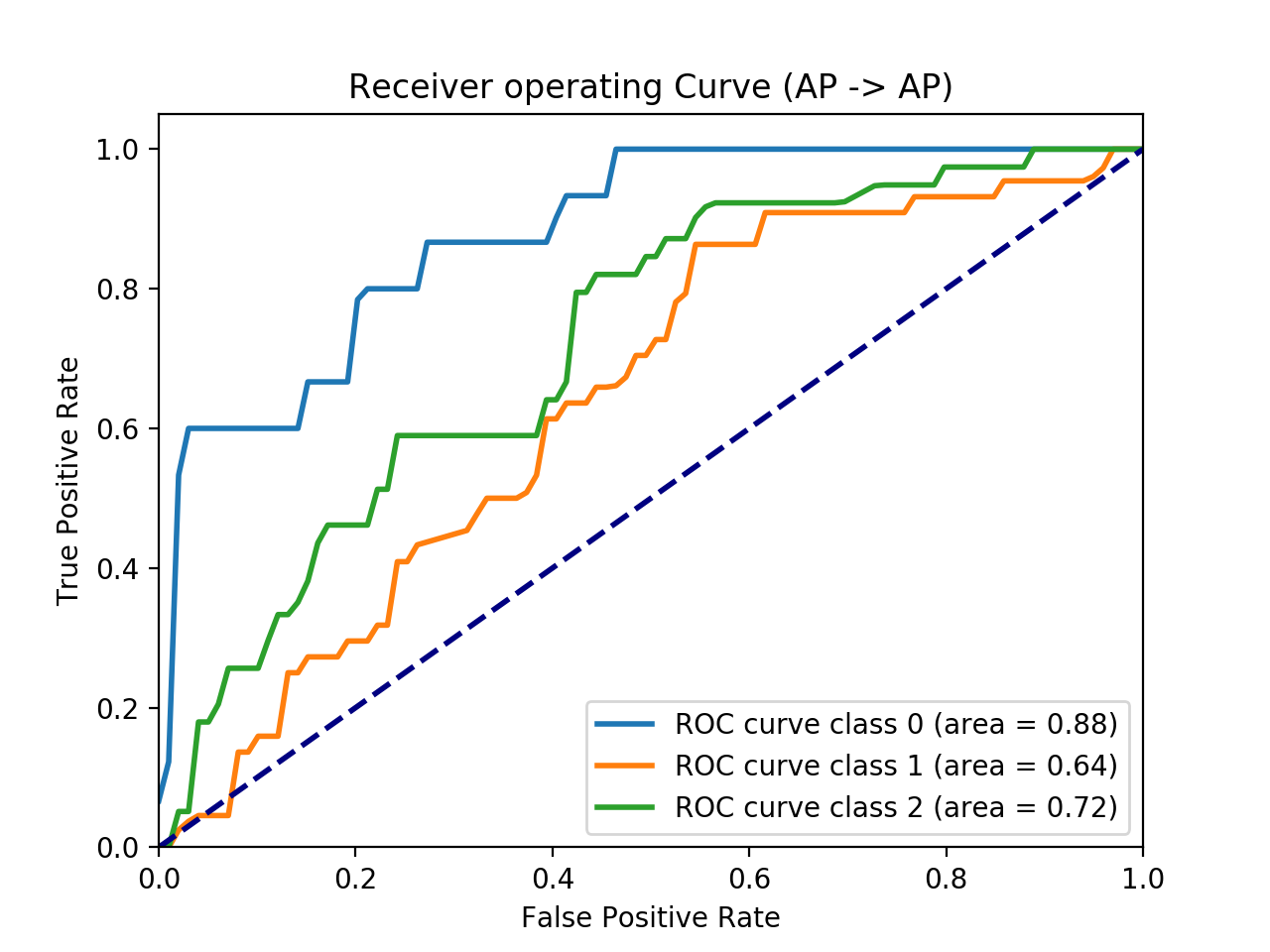}
    \caption{}
    \label{fig:roc1}
    \end{subfigure}
    \begin{subfigure}{0.33\columnwidth}
    \includegraphics[width=\linewidth, height=4cm]{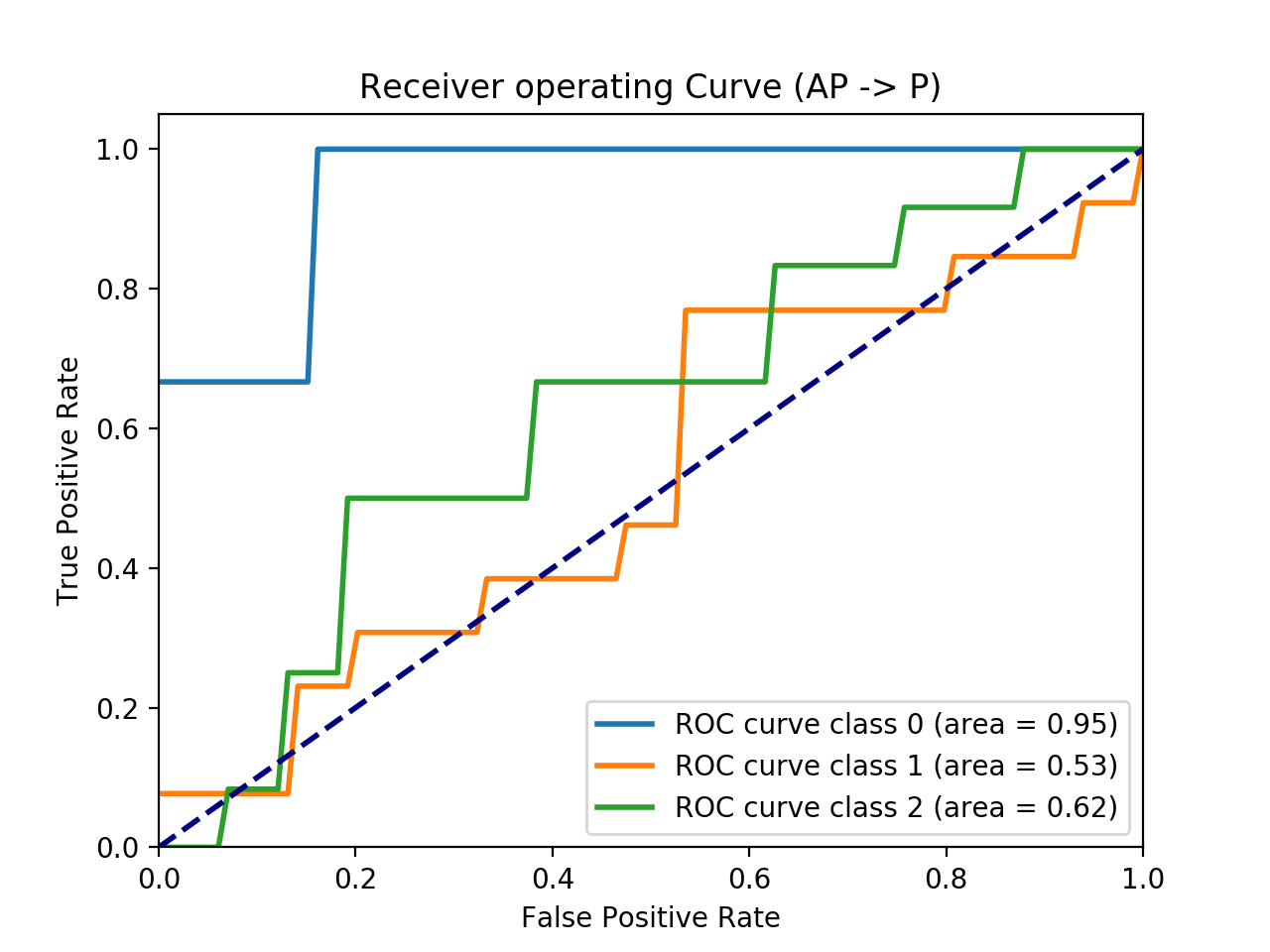}
    \caption{}
    \label{fig:roc2}
    \end{subfigure}
    \begin{subfigure}{0.33\columnwidth}
    \includegraphics[width=\linewidth, height=4cm]{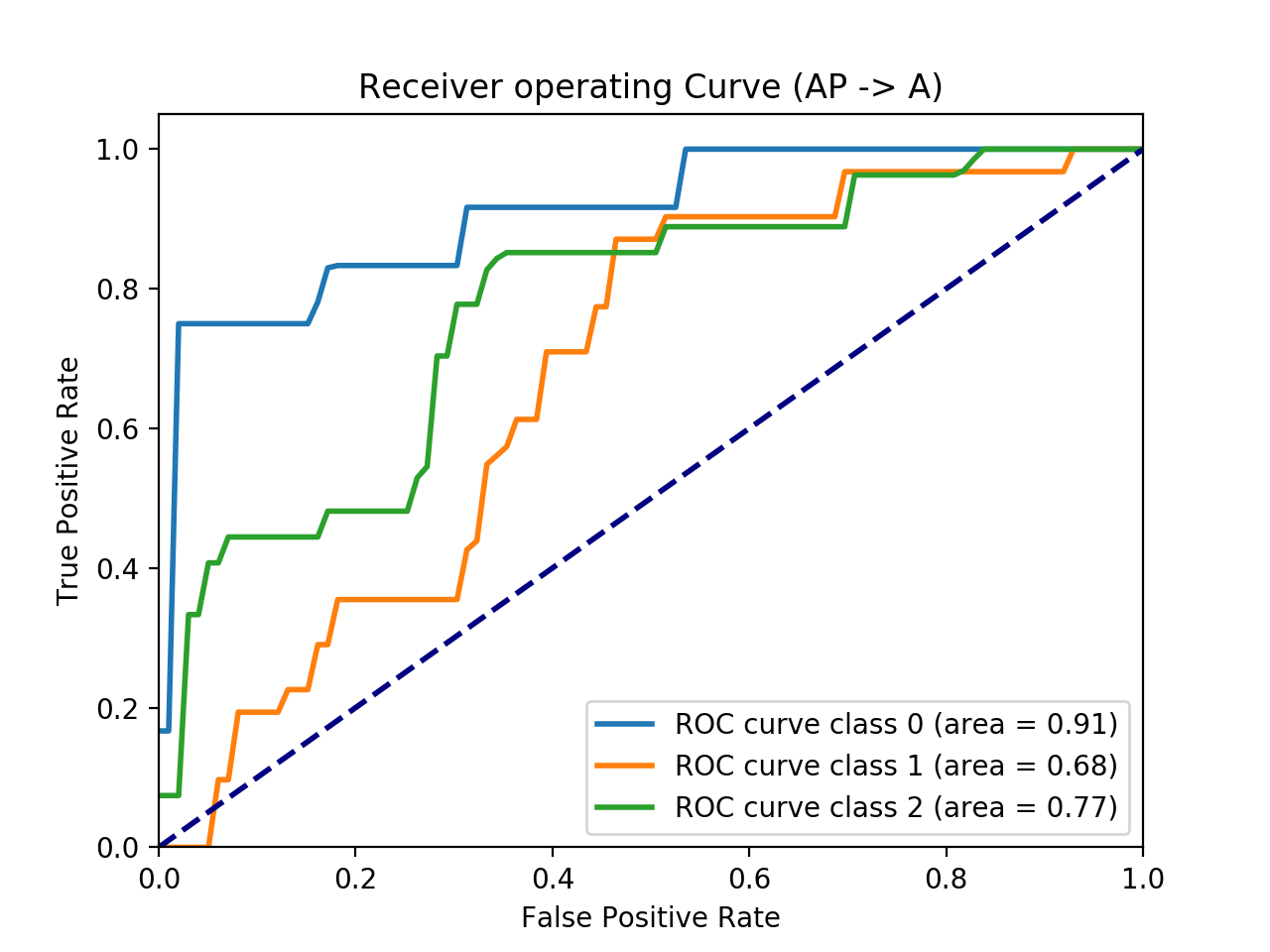}
    \caption{}
    \label{fig:roc3}
    \end{subfigure}
\caption{Receiver Operating Curves (ROC) for three testing scenarios - (a) Train on ADNI and PPMI, Test on ADNI and PPMI, (b) Train on ADNI and PPMI, Test on PPMI, (c) Train on ADNI and PPMI, Test on ADNI - where Class $0$ represents Dementia, class $1$ MCI and class $2$ Normal-CI}
\label{fig:roc}
\end{figure}

Fig~\ref{fig:roc} presents the Receiver Operating Curves (ROC) for three relevant testing scenarios in Table~\ref{tab:results}. Class $0$ represents Dementia, class $1$ MCI and class $2$ Normal-CI. Fig ~\ref{fig:roc1} provides the generalized performance of our model, wherein we train on ADNI$+$PPMI and test on ADNI$+$PPMI. Fig ~\ref{fig:roc2} provides the result most relevant to our use-case. One can observe ~$1.5X$ increase in AUC-Dementia value when we train our model on a combined ADNI$+$PPMI set as compared to PPMI set. This shows drastic gains in prediction of Dementia in Parkinson's patients (PDD) when we used a combined feature space.

\begin{figure}[h!]
    \begin{subfigure}{0.33\columnwidth}
    \includegraphics[width=\linewidth, height=4cm]{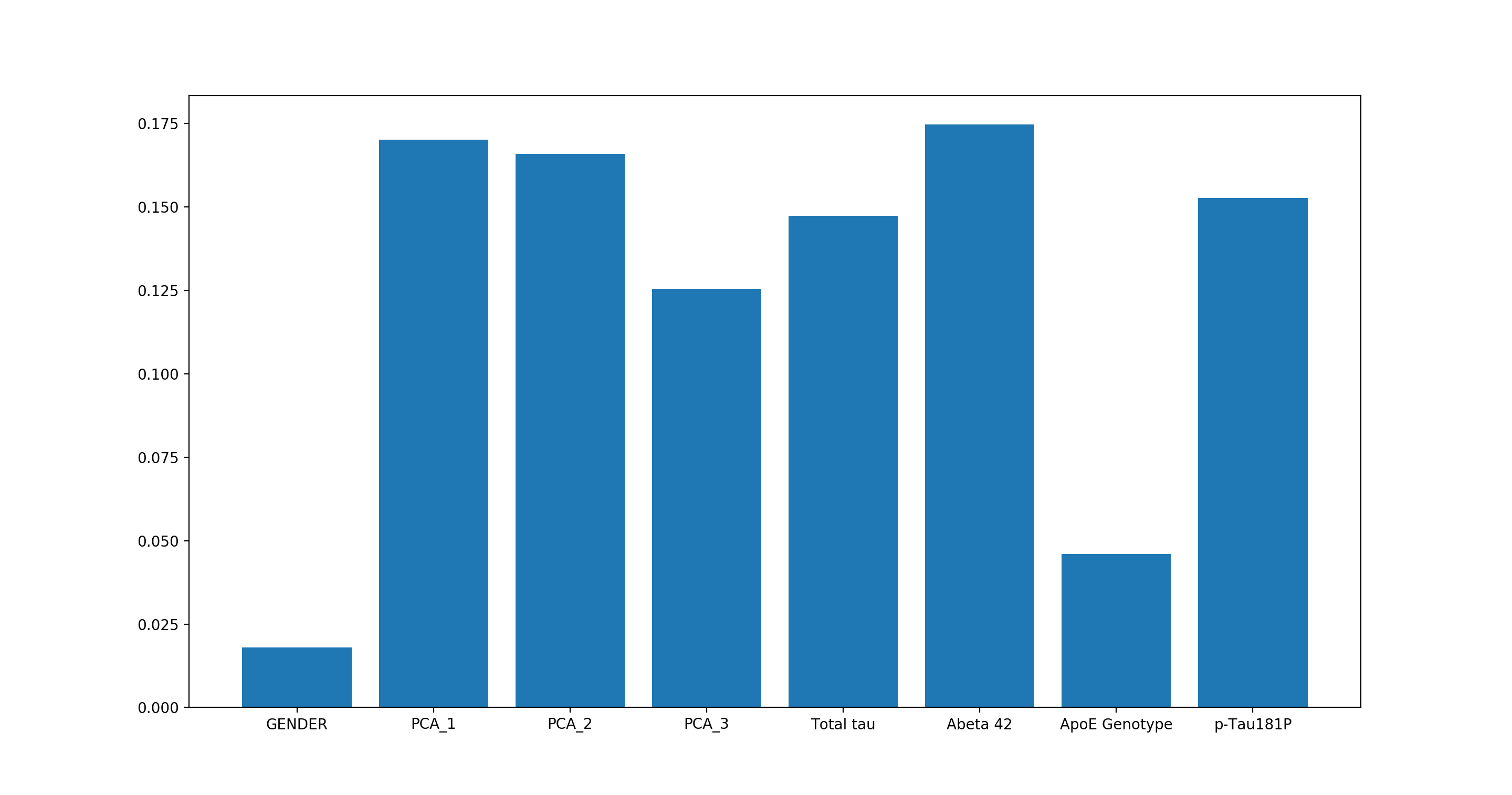} 
    \caption{Feature importance}
    \label{fig:apoe1}
    \end{subfigure}
    \begin{subfigure}{0.33\columnwidth}
    \includegraphics[width=\linewidth, height=4cm]{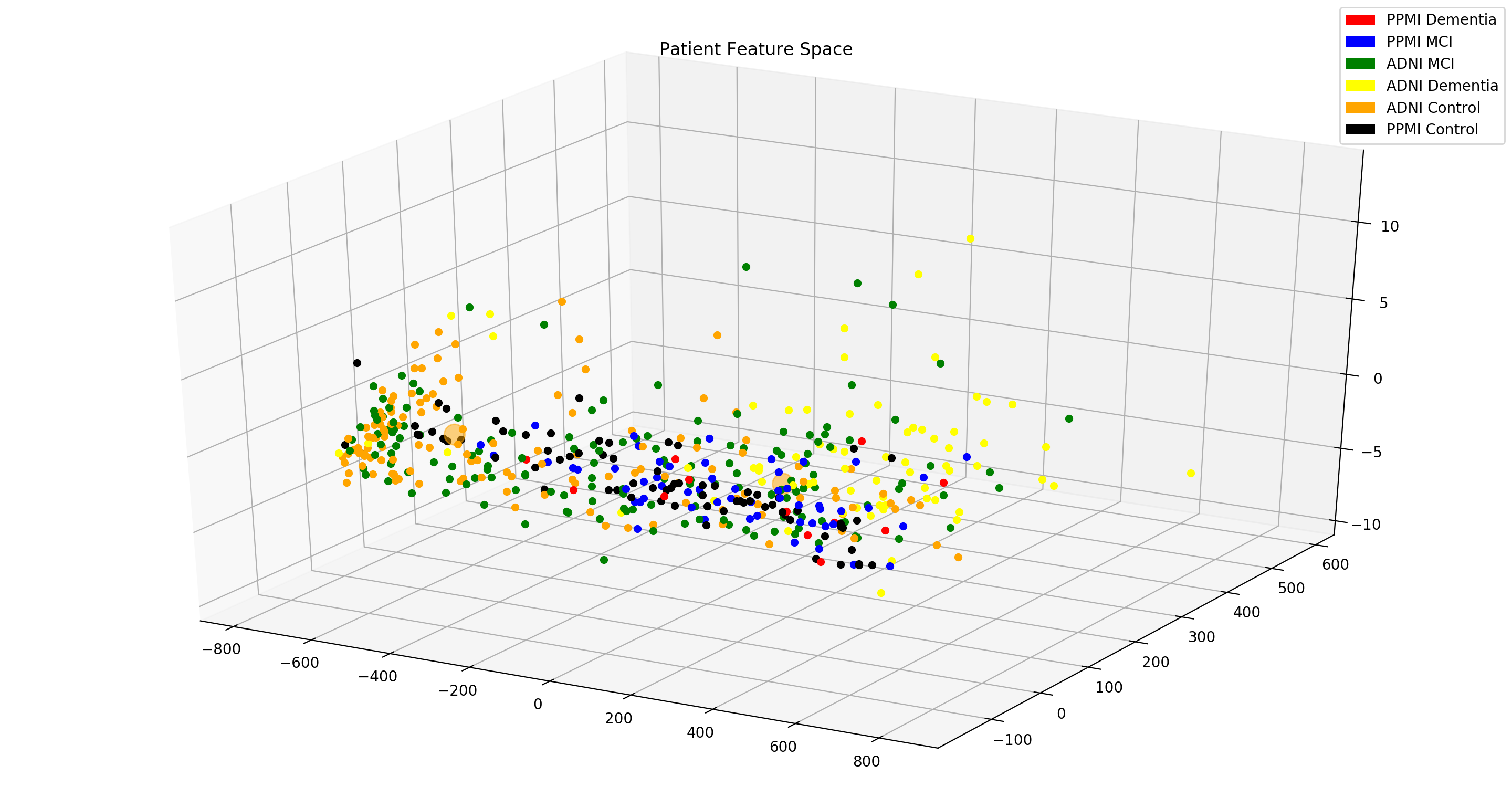} 
    \caption{Feature space without APOE genotype}
    \label{fig:apoe2}
    \end{subfigure}
    \begin{subfigure}{0.33\columnwidth}
    \includegraphics[width=\linewidth, height=4cm]{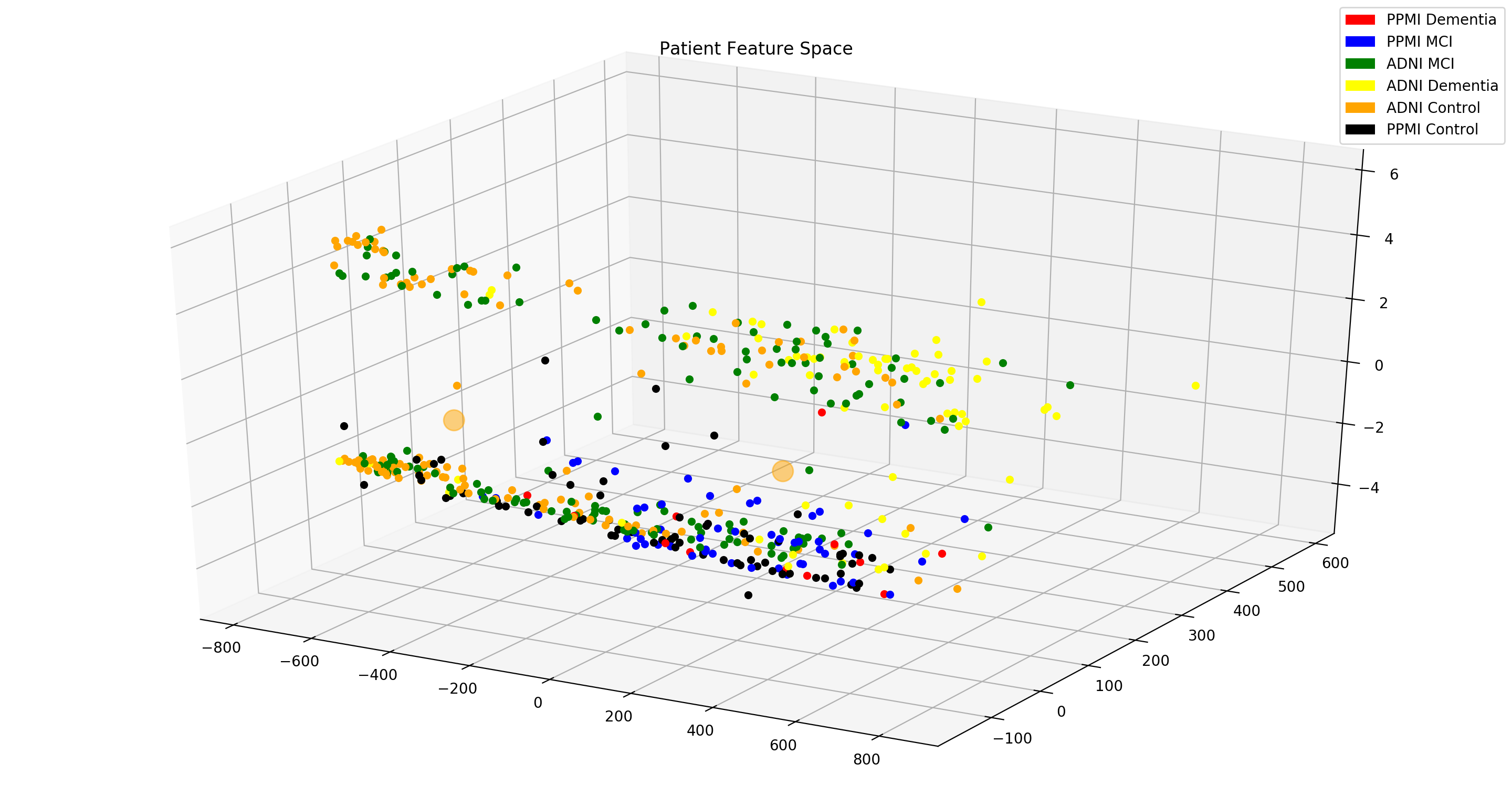}
    \caption{Feature space with APOE genotype}
    \label{fig:apoe3}
    \end{subfigure}
    \label{fig:apoe123}
\end{figure}

Fig ~\ref{fig:apoe1} illustrates importance of each feature used in the classifier. PCA\_1, PCA\_2, PCA\_3 encompass image features. They have a combined importance of ~ 46 $\%$, which points to the necessity of imaging features. Abeta-$42$ has an importance of 17.5 $\%$, which when combined with Total-tau and p-Tau give considerable significance to biospecimen modality. 

Note that ApoE Genotype has a small contribution in prediction outcome, which is counter-intuitive as it has been found to hold high importance when predicting Dementia. We performed exploratory study to clarify this issue. Fig ~\ref{fig:apoe2} shows patient feature space without ApoE Genotype while Fig ~\ref{fig:apoe3} shows feature space with ApoE Genotype. One can clearly observe that ApoE Genotype acts as a discriminating feature when predicting future diagnosis. Fig ~\ref{fig:apoe3} depicts formation of clusters when using all mentioned features, while in Fig ~\ref{fig:apoe2} data has lower visual segregation, which might make classifier less effective. This observation linked with the fact that RF classifier feature importance is not reliable in situations where potential predictor variables vary in their scale of measurement or their number of categories ~\cite{strobl2007bias}, establishes ApoE as an important predictor variable.


Although there is a dearth of research in prediction of Dementia in PD patients (PDD), we were able to use Berlyand et.al ~\cite{berlyand2016} as baseline for our task. They trained a Logistic Regression classifier on a AD cohort of 210 patients to predict control vs Dementia patients. They, then, used the trained model to predict control vs Dementia on a PD cohort of 75 patients with considerably lower performance. Note that we differ in our experiments - (1) usage of raw image features processed through Freesurfer and reduced through PCA, (2) usage of both AD and PD patients for training the model and (3) prediction of Normal\_CI/MCI/Dementia patients instead of Normal/Dementia patients. We use (1) as it opens up our model to better extract information from images and does not impose bias unlike SPARE-AD score. Using (2) exposes our model to a larger data set and (3) provides a more fine-grained outcome.

\section{Conclusion}

This paper makes an attempt at predicting the onset of dementia in patients, with special interest to PD patients. This is a task of high clinical importance. We obtain promising results as compared to the state of the art. We show that using AD patients along with PD patients to build a model to predict the onset of dementia in PD patients is a promising future direction. This also reaffirms our claim that dementia in PD and AD share the same pathology. We believe this can help open up new avenues of research going forward.


\begin{thebibliography}{10}

\bibitem{aarsland2017cognitive}
D.~Aarsland, B.~Creese, M.~Politis, K.~R. Chaudhuri, D.~Weintraub, C.~Ballard,
  et~al.
\newblock Cognitive decline in parkinson disease.
\newblock {\em Nature Reviews Neurology}, 13(4):217, 2017.

\bibitem{berlyand2016}
X.~S. M. I. D. J. R. J.~e. Berlyand~Y, Weintraub~D.
\newblock An alzheimer’s disease-derived biomarker signature identifies
  parkinson’s disease patients with dementia.
\newblock {\em PLoS ONE 11(1)}, 13(4):217, 2017.

\bibitem{compta2011lewy}
Y.~Compta, L.~Parkkinen, S.~S. O'sullivan, J.~Vandrovcova, J.~L. Holton,
  C.~Collins, T.~Lashley, C.~Kallis, D.~R. Williams, R.~de~Silva, et~al.
\newblock Lewy-and alzheimer-type pathologies in parkinson's disease dementia:
  which is more important?
\newblock {\em Brain}, 134(5):1493--1505, 2011.

\bibitem{corder1993gene}
E.~H. Corder, A.~M. Saunders, W.~J. Strittmatter, D.~E. Schmechel, P.~C.
  Gaskell, G.~Small, A.~D. Roses, J.~Haines, and M.~A. Pericak-Vance.
\newblock Gene dose of apolipoprotein e type 4 allele and the risk of
  alzheimer's disease in late onset families.
\newblock {\em Science}, 261(5123):921--923, 1993.

\bibitem{Dalrymple-Alford1717}
J.~Dalrymple-Alford, M.~MacAskill, C.~Nakas, L.~Livingston, C.~Graham,
  G.~Crucian, T.~Melzer, J.~Kirwan, R.~Keenan, S.~Wells, R.~Porter, R.~Watts,
  and T.~Anderson.
\newblock The moca.
\newblock {\em Neurology}, 75(19):1717--1725, 2010.

\bibitem{fischl2012freesurfer}
B.~Fischl.
\newblock Freesurfer.
\newblock {\em Neuroimage}, 62(2):774--781, 2012.

\bibitem{Hoops1738}
S.~Hoops, S.~Nazem, A.~D. Siderowf, J.~E. Duda, S.~X. Xie, M.~B. Stern, and
  D.~Weintraub.
\newblock Validity of the moca and mmse in the detection of mci and dementia in
  parkinson disease.
\newblock {\em Neurology}, 73(21):1738--1745, 2009.

\bibitem{Irvine2008}
S.~G.~o. Irvine~GB, El-Agnaf~OM.
\newblock Protein aggregation in the brain: The molecular basis for
  alzheimer’s and parkinson’s diseases. molecular medicine.
\newblock {\em Molecular Medicine}, 14(7-8):451--464, 2008.

\bibitem{irwin2017neuropathological}
D.~J. Irwin, M.~Grossman, D.~Weintraub, H.~I. Hurtig, J.~E. Duda, S.~X. Xie,
  E.~B. Lee, V.~M. Van~Deerlin, O.~L. Lopez, J.~K. Kofler, et~al.
\newblock Neuropathological and genetic correlates of survival and dementia
  onset in synucleinopathies: a retrospective analysis.
\newblock {\em The Lancet Neurology}, 16(1):55--65, 2017.

\bibitem{kang2013association}
J.-H. Kang, D.~J. Irwin, A.~S. Chen-Plotkin, A.~Siderowf, C.~Caspell, C.~S.
  Coffey, T.~Walig{\'o}rska, P.~Taylor, S.~Pan, M.~Frasier, et~al.
\newblock Association of cerebrospinal fluid $\beta$-amyloid 1-42, t-tau,
  p-tau181, and $\alpha$-synuclein levels with clinical features of drug-naive
  patients with early parkinson disease.
\newblock {\em JAMA neurology}, 70(10):1277--1287, 2013.

\bibitem{kempster2010relationships}
P.~A. Kempster, S.~S. O’Sullivan, J.~L. Holton, T.~Revesz, and A.~J. Lees.
\newblock Relationships between age and late progression of parkinson’s
  disease: a clinico-pathological study.
\newblock {\em Brain}, 133(6):1755--1762, 2010.

\bibitem{mathotaarachchi2017identifying}
S.~Mathotaarachchi, T.~A. Pascoal, M.~Shin, A.~L. Benedet, M.~S. Kang,
  T.~Beaudry, V.~S. Fonov, S.~Gauthier, and P.~Rosa-Neto.
\newblock Identifying incipient dementia individuals using machine learning and
  amyloid imaging.
\newblock {\em Neurobiology of aging}, 59:80--90, 2017.

\bibitem{nasreddine2005montreal}
Z.~S. Nasreddine, N.~A. Phillips, V.~B{\'e}dirian, S.~Charbonneau,
  V.~Whitehead, I.~Collin, J.~L. Cummings, and H.~Chertkow.
\newblock The montreal cognitive assessment, moca: a brief screening tool for
  mild cognitive impairment.
\newblock {\em Journal of the American Geriatrics Society}, 53(4):695--699,
  2005.

\bibitem{pearson1901liii}
K.~Pearson.
\newblock Liii. on lines and planes of closest fit to systems of points in
  space.
\newblock {\em The London, Edinburgh, and Dublin Philosophical Magazine and
  Journal of Science}, 2(11):559--572, 1901.

\bibitem{petersen2002mild}
R.~C. Petersen.
\newblock Mild cognitive impairment: transition from aging to alzheimer's
  disease.
\newblock {\em Alzheimer's disease: advances in etiology, pathogenesis and
  therapeutics}, pages 141--151, 2002.

\bibitem{polvikoski1995apolipoprotein}
T.~Polvikoski, R.~Sulkava, M.~Haltia, K.~Kainulainen, A.~Vuorio,
  A.~Verkkoniemi, L.~Niinist{\"o}, P.~Halonen, and K.~Kontula.
\newblock Apolipoprotein e, dementia, and cortical deposition of
  $\beta$-amyloid protein.
\newblock {\em New England Journal of Medicine}, 333(19):1242--1248, 1995.

\bibitem{sadigh2012association}
S.~Sadigh-Eteghad, M.~Talebi, and M.~Farhoudi.
\newblock Association of apolipoprotein e epsilon 4 allele with sporadic late
  onset alzheimer’s disease.
\newblock {\em A meta-analysis. Neurosciences (Riyadh)}, 17(4):321--326, 2012.

\bibitem{shaw2009cerebrospinal}
L.~M. Shaw, H.~Vanderstichele, M.~Knapik-Czajka, C.~M. Clark, P.~S. Aisen,
  R.~C. Petersen, K.~Blennow, H.~Soares, A.~Simon, P.~Lewczuk, et~al.
\newblock Cerebrospinal fluid biomarker signature in alzheimer's disease
  neuroimaging initiative subjects.
\newblock {\em Annals of neurology}, 65(4):403--413, 2009.

\bibitem{strobl2007bias}
C.~Strobl, A.-L. Boulesteix, A.~Zeileis, and T.~Hothorn.
\newblock Bias in random forest variable importance measures: Illustrations,
  sources and a solution.
\newblock {\em BMC bioinformatics}, 8(1):25, 2007.

\bibitem{thomann2009association}
P.~A. Thomann, E.~Kaiser, P.~Sch{\"o}nknecht, J.~Pantel, M.~Essig, and
  J.~Schr{\"o}der.
\newblock Association of total tau and phosphorylated tau 181 protein levels in
  cerebrospinal fluid with cerebral atrophy in mild cognitive impairment and
  alzheimer disease.
\newblock {\em Journal of psychiatry \& neuroscience: JPN}, 34(2):136, 2009.

\bibitem{weintraub2011alzheimer}
D.~Weintraub, N.~Dietz, J.~E. Duda, D.~A. Wolk, J.~Doshi, S.~X. Xie,
  C.~Davatzikos, C.~M. Clark, and A.~Siderowf.
\newblock Alzheimer's disease pattern of brain atrophy predicts cognitive
  decline in parkinson's disease.
\newblock {\em Brain}, 135(1):170--180, 2011.

\bibitem{younkin1998role}
S.~G. Younkin.
\newblock The role of a$\beta$42 in alzheimer's disease.
\newblock {\em Journal of Physiology-Paris}, 92(3-4):289--292, 1998.

\end{thebibliography}

\section{Appendix}

\begin{table}[h!]
\centering
\caption{Statistical summary of patients with assigned labels.}
\label{tab:1}
\begin{tabular}{ ccccc } 
\hline
Dataset & Baseline Diagnosis & Future Diagnosis & Number of patients \\
\hline
\multirow{6}{4em}{ADNI} & Normal CI & Normal CI & 131\\ 
& Normal CI & MCI & 11\\ 
& Normal CI & Dementia & 2\\
& MCI & MCI & 144\\
& MCI & Dementia & 48\\
& Dementia & Dementia & 19\\
\hline
\multirow{6}{4em}{PPMI} & Normal CI & Normal CI & 75\\ 
& Normal CI & MCI & 35\\ 
& Normal CI & Dementia & 4\\
& MCI & MCI & 20\\
& MCI & Dementia & 7\\
& Dementia & Dementia & 1\\
\hline
\multirow{6}{4em}{Combined} & Normal CI & Normal CI & 206\\ 
& Normal CI & MCI & 46\\ 
& Normal CI & Dementia & 6\\
& MCI & MCI & 164\\
& MCI & Dementia & 55\\
& Dementia & Dementia & 29\\
\hline
\end{tabular}
\end{table}

Table ~\ref{tab:1} summarizes the composition of different types of progressing patients used in our prediction task. Note that we are only concerned with disease progression; hence, we do not consider disease reverter cases (patients who improve in disease condition).

The eligibility criteria of MCI were participants with minimental state examination (MMSE) scores equal to or greater than 24, a clinical dementia rating of 0.5, subjective and objective memory loss, normal activities of daily living, and the absence of other neuropsychiatric disorders, including Alzheimer’s disease, at baseline \cite{petersen2002mild}.

\end{document}